\documentclass[twocolumn,superscriptaddress,showpacs,aps,amsmath,amssymb,prl]{revtex4}
\usepackage{amsfonts}
\usepackage{amsmath}
\usepackage{amssymb}
\usepackage{bm}
\usepackage{graphicx}

\newcommand{\bea}{\begin{eqnarray}}
\newcommand{\eea}{\end{eqnarray}}

\begin{document}

\title{Tracking the electronic oscillation in molecule with tunneling microscopy}

\author{Rulin Wang}  \email{rulin11@qdu.edu.cn}
\affiliation{College of Physics, Qingdao University, No. 308 Ningxia Road,
Qingdao, 266071, China}
\author{Fuzhen Bi}
\affiliation{Qingdao Institute of Bioenergy and Bioprocess Technology, Chinese Academy of Sciences,
Qingdao, 266101, China}

\author{Wencai Lu}
\affiliation{College of Physics, Qingdao University, No. 308 Ningxia Road,
Qingdao, 266071, China}
\author{Xiao Zheng} 
\affiliation{Hefei National Laboratory for Physical Sciences at the
Microscale, University of Science and Technology of China, Hefei, Anhui
230026, China}
\author{ChiYung Yam}  \email{yamcy@csrc.ac.cn}
\affiliation{Beijing Computational Science Research Center, Haidian District, Beijing, 100193, China}

\date{\today}



%
\begin{abstract}
Visualizing and controlling electron dynamics over femtosecond timescale play a key role in the design of next-generation electronic device.  
Using simulations, we demonstrate the electronic oscillation inside naphthalene molecule can be tracked by means of the tuning of delay time between two identical femtosecond laser pulses.
Both the frequency and decay time of the oscillation are detected by the tunneling charge through the junction of scanning tunneling microscopy. 
And the tunneling charge is sensitive to the carrier-envelope phase (CEP) for few-cycle long optical pulses.  
While this sensitivity to CEP will disappear with the increase of time-length of pulses. 
Our simulation results show that it is possible to visualize and control the electron dynamics inside molecule by one or two femtosecond laser pulses. 
 
\end{abstract}

\maketitle

%
Scanning tunneling microscope (STM) is firstly proposed to image static surface patterns in atomic-scale with steady tunneling current varying by the distance between tip and surface.\cite{Bin1982Tunn,Bin1982Surf}
After that, many efforts are devoted to probe the ultrafast dynamic phenomenon on atomic time-scales.\cite{Nunes1993Pico,Wu2006Atomic,Wu2010Two,Tera2010Real,Cocker2013Ultra,Yosh2014Prob,Kwok2019Stm}
In the experiment, this is usually realized with isolated photon pulses focused onto STM junction and the transient bias voltages created by pulses lead to the transient currents through the junction.\cite{Tera2010Real,Kwok2019Stm}
The dynamic phenomenon is detected with time-integration of the transient tunneling currents (tunneling charge). 
Terahertz STM (THz-STM) technique\cite{Cocker2013Ultra,Jel2017Ultra} has been developed to track the real-time dynamics of molecular vibrations \cite{Cock2016Track} and single-molecule structural transition \cite{Li2017Joint} in picosecond time-scale.    
Recently, the dynamics of electronic oscillations inside a Au nanorod has been visualized with few-femtosecond long optical pulses.\cite{Garg2020Atto}  
While the electronic dynamics in molecule is very importance for chemical reaction, structural transition and practical application. 
In our research, we focus on visualizing and controlling electronic dynamics in molecule. 
For the isolated optical pulses, especially few-cycle long pulses, the carrier-envelope phase(CEP) is very important to control the tunneling charge.\cite{Rybka2016Sub,Yoshi2016Real} 
The change of CEP can modulate the transient bias voltage around the STM nanotip, and consequently change the tunneling currents through the junction.       
The geometric confinement of incident pulses by nanostructure of STM tip can strongly enhance the transient bias voltage, with strength of several volts across the nanotip.\cite{Garg2020Atto,Muller2020Phase}
This large strength of voltages can create observable light-field-driven currents in the experiments. 
The CEP of pulses have been used to control the number and direction of tunneling charge from the STM tip to sample or vice versa.\cite{Yoshi2016Real,Yoshi2018Tail} 
Similar phenomenon of CEP controllable currents has also been reported in pulses irradiated metal-semiconductor-metal nanojunctions.\cite{Kuri1989Phase,Dupont1995Phase,Schif2013Opti,Chen2018Stark}
Most researches in this area focus on the dependence of tunneling currents on central frequency and energies of the laser pulses. 
Whether and how this CEP sensitivity of tunneling charge disappearing with increase of time-length of pulses is still unclear. 
In this paper, using simulation, we will show the variation of tunneling charge with the time-length of isolated pulses, and how to track the electronic dynamics in a molecule with STM.  
The dynamic process is described with time-dependent functional theory for open system(TDDFT-OS). \cite{Burke2005Time,Kurth2005Time,Zheng2007Time,Zheng2010Time,Kwok2013Time,Chen2018Time}
The whole system is divided into two parts: the part around the STM nanotip as open system; the other part as environment. 
The environment is assumed in thermal equilibrium state, and provides the dissipative paths for the open system. 
In nonequilibrium Green's function (NEGF) framework, self-energies represents the effects of environment on open system and Fermi-Dirac distribution gives the electron occupation in environment. 
To improve computational efficiency, Pad\'{e} and Lorentzian decomposition(PLD) is proposed to decompose Fermi-Dirac distribution and self-energies, respectively. And the numbers of Pad\'{e} functions and Lorentzian functions determine the simulation efficiency directly. (see supplementary text section I for detail)   
TDDFT-NEGF had been used to simulate quasi-one-dimensional electronic device. Recently, we develop numerical scheme to obtain the self-energies for monolayer surface \cite{Wang2013Time,Wang2015Time} and bulk materials \cite{Wang2019Theo}, and simulate the transient electronic dynamics in quasi-two-dimensional and three-dimensional systems.  
The interesting thing is that computational cost for bulk material is least due to the smooth density of states of the bulk material.\cite{Wang2019Theo}  
In this way, we can simulate the electronic dynamic process in STM system with bulk substrate.

\begin{figure}[t]
  \includegraphics[width=1.0\columnwidth]{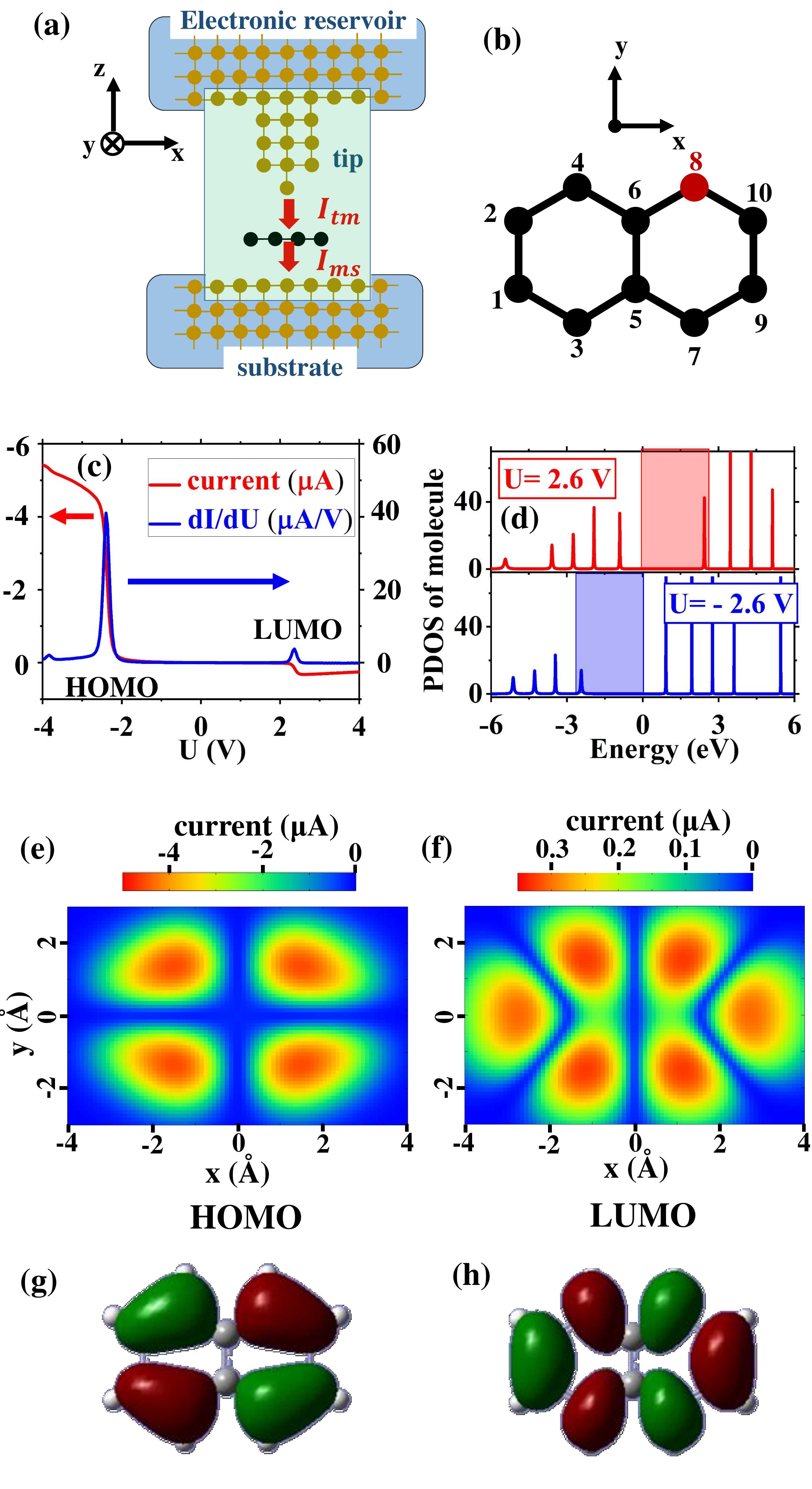}
  \caption{(a) Schematic  of modeling STM with a naphthalene molecule. (b) The structure of naphthalene molecule. (c) Steady currents (red line) and differential conductance (blue line) varying with bias voltage $U$. (d) The projected density of states (PDOS) of the molecule under the voltage $U = 2.6 V$ and $U = -2.6 V$. Simulated constant height STM image of the HOMO (e) and LUMO (f) under the voltage $U = -2.6V$ and $U = 2.6V$, respectively. The spatial distribution of the HOMO (g) and LUMO (h) from DFT simulation.}
  \label{FigStruc}
\end{figure}

Here, we simulate a modeling STM system with a naphthalene molecule around the junction, as shown in Fig.\ref{FigStruc}a. 
In our calculation, the main role of STM (including nanotip and substrate) is to provide the dissipative paths for the tunneling currents through the molecule.  
For simplicity, the structure of STM system are assumed to be simple cubic lattice. 
And the naphthalene molecule includes ten carbon atoms with the structure shown in Fig.\ref{FigStruc}b. 
The bond lengths inside the STM system and molecule are set to $l_0=1 \AA$ and $l_m=1.42 \AA$, respectively.  
The whole system is described by a tight-binding Hamiltonian $\bm H$, which consists of three parts: STM Hamiltonian; molecule Hamiltonian and interaction Hamiltonian between STM and molecule. 
Each atom has a single level and on-site energy is $\epsilon = 0$. The nearest-neighbour coupling inside the STM system is $\gamma_0 = -2.0 eV$ and the coupling inside molecule is assumed to be the same value as strength in graphene $\gamma_m = -2.7eV$.\cite{Neto2009Elec} 
The interaction Hamiltonian $\bm H_{int}$ depends on the atomic distance $r$ between STM and molecule: $\bm H_{int}(r) = \gamma_t e^{-\beta(r-l_t)/ l_t}$,\cite{Mor2010Flat} where $\gamma_t = (\gamma_0+\gamma_m)/2$ and $l_t = (l_0+l_m)/2$. The parameter $\beta$ is set to $3$ in our calculation. 
The molecule is separated from the substrate by $2.5 \AA$ to avoid the strong interaction and maintain the intrinsic electron structure of molecule. 
In the experiment, this electronic decoupling of molecule from substrate can be realized by inserting a thin insulating layer. 
The TDDFT-NEGF approach is employed to simulate the electron dynamical process and whole system is divided into two parts: the green area in Fig.\ref{FigStruc}a as open system; the other part as environment.  
The open system includes 108 atoms. 
The effects of environment on open system is described by self energies, which is obtained with the $k$-sampling renormalization approach.\cite{Wang2019Theo}
The super cell for calculating the self energies consists of $35(7 \times 5 \times 1)$ atoms and a set of $(15 \times 15)$ $k$-points on $x$-$y$ plane is used.
In the simulation, we use the PLD scheme to reduce the computational cost. The numbers of Pad{\'e} functions and Lorentzian functions are set to $N_p = 20$ and $N_l = 10$ to guarantee the calculation precision (for details, see Figure S1 in supporting information).     
The system temperature is $T=300K$.

We first carry out simulation of steady states in the STM system. 
The nanotip is on the top of the $8$-$th$ atom of molecule. The tunneling gap between nanotip and molecule is set to $2 \AA$.  
The red line in Fig\ref{FigStruc}c shows the simulation results of steady currents varying with the bias voltage $U$ applied on the nanotip.   
The positive currents mean the charge tunneling from nanotip to substrate and the negative values mean the opposite direction.   
For the low bias voltage, there is nearly no current due to lack of molecular state inside the bias window. 
When the voltage $U$ increases to about $-2.6 V$, the corresponding steady current increases sharply because the highest occupied molecular orbital (HOMO) is included into the bias voltage window, as shown in Fig.\ref{FigStruc}d.  
Similarly, the increase of current around $U = 2.6 V$ is due to the tunneling through lowest unoccupied molecular orbital(LUMO).   
This contribution of currents is more clearer in the simulation results of differential conductance (blue line in Fig.\ref{FigStruc}c), two peaks are corresponding to HOMO and LUMO respectively.   
The negative-voltage induced currents are much larger than the positive-voltage induced currents. 
To investigate the spatial distribution of molecular orbitals, we move the nanotip on the $x$-$y$ plane with constant tunneling gap width $2 \AA$ and calculate the steady current for each nanotip position. 
Fig.\ref{FigStruc}e (Fig.\ref{FigStruc}f) plots the constant-height image of naphthalene molecule under the bias voltage $U=-2.6 V$ ($U=2.6V$), and depicts the geometry shape of HOMO (LUMO).    
For easy comparison, Fig.\ref{FigStruc}g and Fig.\ref{FigStruc}h show the spatial distribution of HOMO and LUMO for the free naphthalene molecule determined from density functional theory (DFT) simulations.  
We can see the STM image of orbitals calculated from our model are in good agreement with the simulation results of DFT. 

\begin{figure}[t]
  \includegraphics[width=1.0\columnwidth]{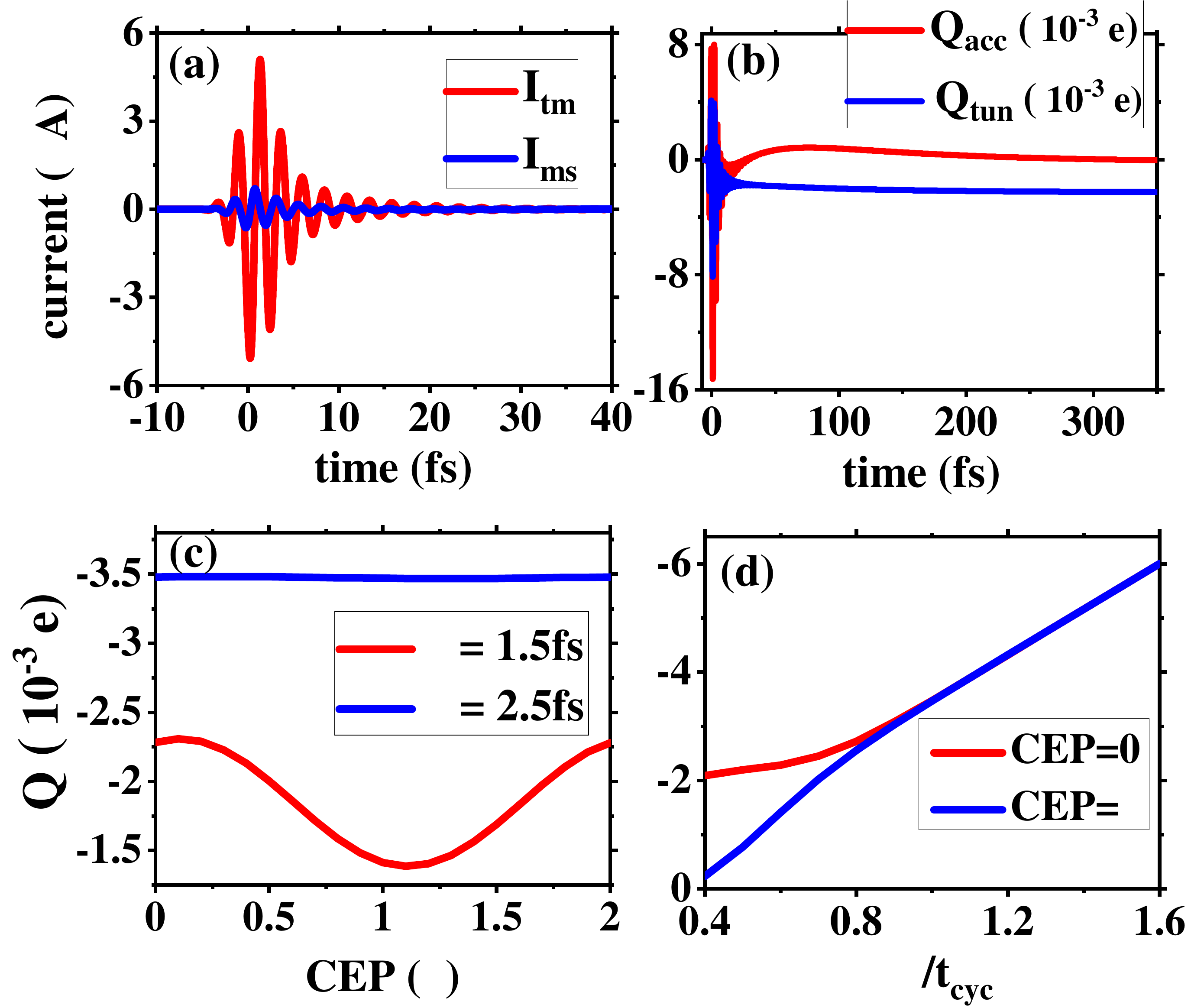}
  \caption{(a) The transient current $I_{tm}$ and $I_{ms}$ driven by a laser pulse with ($\varphi=0,\sigma =1.5fs$). (b) The tunneling charge through molecule $Q_{tun}$ and accumulation charge inside molecule $Q_{acc}$ varying with time. The total tunneling charge $Q$ as functions of the CEP (c) and time-length (d) of laser pulses.} 
  \label{PhaseCurrent}
\end{figure}

We then move to the simulation of transient states induced by the femtosecond laser pulses. 
The laser pulses are considered as linearly polarized light and the polarization axis is along the nanotip.   
The laser electric field leads to the transient bias voltage $U_{pul}(t)$ across STM junction. 
In the simulation, we assume the transient voltage as $U_{pul}(t) = U_0 G(t) sin [2\pi(t-t_0)/t_{cyc} + \varphi]$. 
Here $t_0$, $t_{cyc}$ and $\varphi$ denote the center time, cycle time and CEP of laser pulse, respectively. 
A Gaussian function $G(t) = e^{-(t-t_0)^2/2\sigma^2}$ is employed to regulate time-length of pulses with the width $\sigma$. 
And $U_0$ is amplitude of transient voltage, which can be strongly enhanced to several volts by geometric confinement of laser pulses around the STM nanotip.\cite{Garg2020Atto,Muller2020Phase}
At $t_0 = 0 fs$, a laser pulse $(U_0 = -2.6V, t_{cyc} = 2.5 fs, \varphi =0, \sigma = 1.5 fs)$ is incident on the STM junction.    
The transient currents from nanotip to molecule $I_{tm}$ and from molecule to substrate $I_{ms}$ are shown in Fig.\ref{PhaseCurrent}a. 
The fluctuation of current $I_{tm}$ is larger than the current $I_{ms}$, due to the stronger coupling strength between nanotip and molecule. 
While this kind of fluctuation in femtosecond scale, is so fast that the directly measurement in experiment is impossible.   
Thus, time-integration of currents (tunneling charge) is of interest in investigation of transient dynamical process. 
In our calculation, this tunneling charge is represented by: $Q_{tun} (t) = \int_{-\infty}^t d\tau (I_{tm}+I_{ms})/2$ and the accumulation charge inside the molecule is: $Q_{acc}(t) = \int_{-\infty}^t d\tau (I_{tm}-I_{ms})$.  
From simulation result of $Q_{acc}(t)$, the red line in Fig.\ref{PhaseCurrent}b, we can see the non-equilibrium state driven by laser pulse will tend to return the initial equilibrium state as time goes on. 
And at $t = 250 fs$, the effect of pulse is totally removed from the molecule.
In the mean time, the tunneling charge reaches a stabilized value, which can be measured directly in experiment.    
Hereafter, we will only consider this total tunneling charge $Q$ response to incident laser pulses.  
Figure\ref{PhaseCurrent}c plots the tunneling charge $Q$ varying with the CEP of pulses. 
When $\sigma = 1.5 fs$, the tunneling charge is sensitivity to the CEP with the maximum value $Q(\varphi=0)$ and minimum value $Q(\varphi=\pi)$. 
This phase-controllable tunneling charge has been observed in various experiment. \cite{Schif2013Opti,Rybka2016Sub,Yoshi2016Real,Yoshi2018Tail,Garg2020Atto}
The change of CEP would modulate the negative and positive proportion of transient voltage, and the current driven by negative voltage is much lager than the positive voltage(as shown in Fig.\ref{FigStruc}c).  
Thus, we obtain the maximum and minimum value of tunneling charge when negative proportion of voltage reaches its extreme at $\varphi=0$ and $\varphi = \pi$. 
With increase of time-length of laser pulse, the asymmetry of transient voltage will gradually vanish, the tunneling charge for $\varphi=0$ and $\varphi=\pi$ would tend to the same value as shown in Fig.\ref{PhaseCurrent}d.    
At $\sigma = 2.5fs$ $(\sigma/t_{cyc} = 1)$, the CEP dependence is eliminated and we get the same value of tunneling charge $Q = -0.0035 e$ for various $\varphi$ (blue line in Fig.\ref{PhaseCurrent}c).  
When $\sigma > 2.5 fs$, the linear growth of $Q$ is relate to the increase of cycle number of pulses with time-length $\sigma$.    
The slope of growth is determined by the difference of the tunneling charge induced by the negative and positive voltage for each cycle of pulses.  
Similar results are obtained if we move nanotip to the different positions on molecule (see Figure S2 in supporting information).

\begin{figure}[htb]
  \includegraphics[width=1.0\columnwidth]{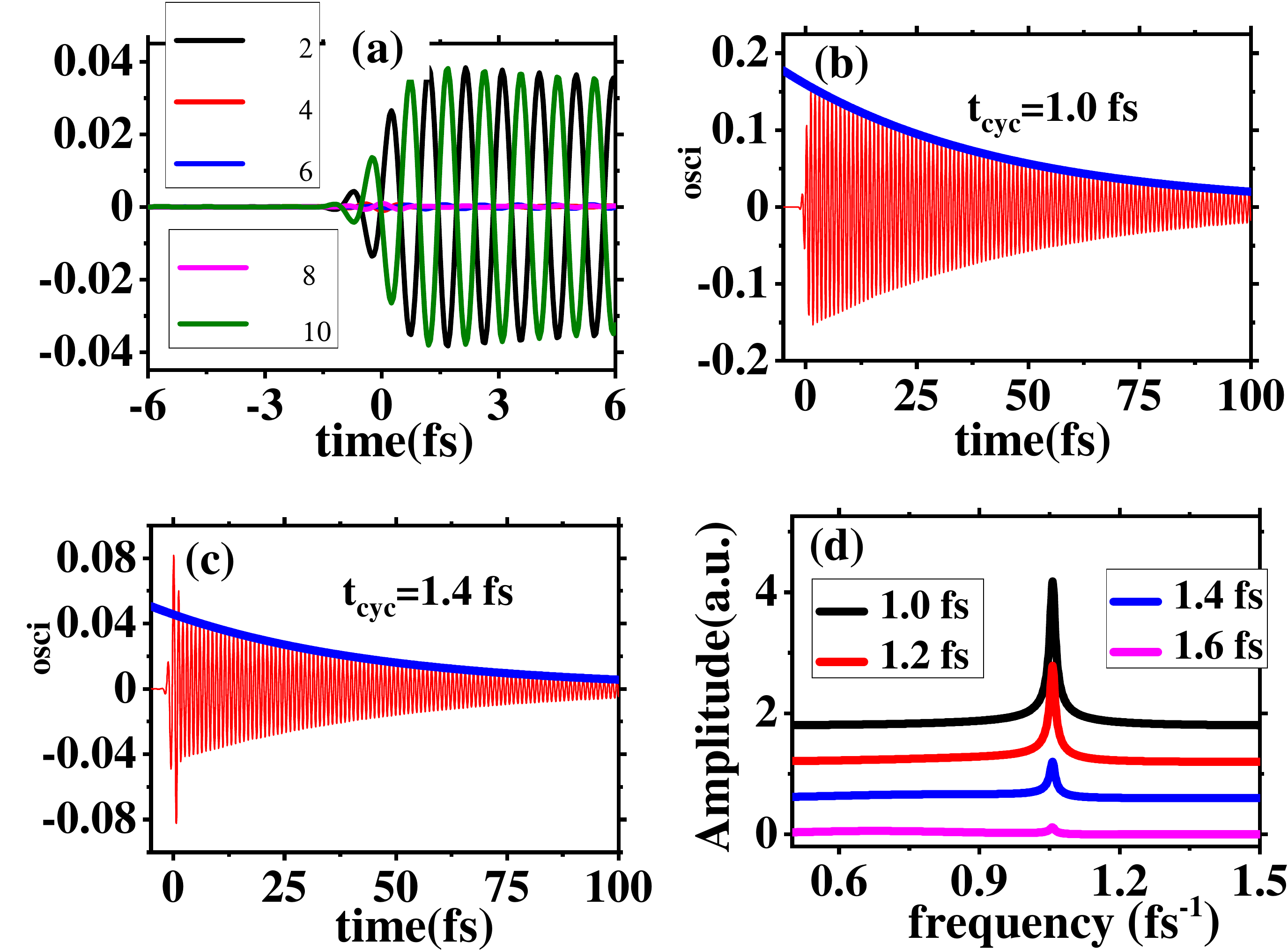}
  \caption{(a) The change of electron density on the $2$-$nd$, $4$-$th$, $6$-$th$, $8$-$th$, $10$-$th$ atoms of naphthalene molecule stimulated by a laser pulses ($U_0  = 0.2 V/nm, t_{cyc} = 1.0 fs, \varphi = 0, \sigma = 0.6 t_{cyc}$). The difference of electron density at two sides of molecule $\Delta \rho_{osci}$ excited by laser pulses $t_{cyc} = 1.0 fs$ (b) and $t_{cyc} = 1.4 fs$ (c). (d) The Fourier transform of $\Delta \rho_{osci}$ for the laser pulses $t_{cyc} = 1.0,1.2,1.4,1.6 fs$.}
  \label{Osci}
\end{figure}

We now investigate how to track the electronic dynamics inside naphthalene molecule. 
And the electronic dynamics is stimulated by laser pulses polarized on molecular plane and along the $x$-axis.  
The parameters of pulse are set to $U_0  = 0.2 V/nm, t_{cyc} = 1.0 fs, \varphi = 0, \sigma = 0.6 t_{cyc}$.  
The transient electric field will drive the molecule out of the equilibrium state.  
Fig.\ref{Osci}a plots the time-varying change of electronic density on $i$-$th$ atom of molecule: $\Delta \rho_{i} = \rho_{i} - \rho_{i}^{equ}$.
It is clear that the fluctuation of electronic density is mainly occurred on the ends of molecule ($2$-$th$ atom,$10$-$th$ atom), and establishes electronic oscillation in $x$-axis with opposite value of $\Delta \rho_{i}$ at two sides of molecule.     
The electronic oscillation is described by the difference of electronic density at two sides: $\Delta \rho_{osci} = (\Delta \rho_{1} + \Delta \rho_{2} - \Delta \rho_{9} - \Delta \rho_{10})$, as shown in Fig.\ref{Osci}b. 
We can see the oscillation will decay after excitation by laser pulse at $t=0$.
This is because the weak interaction between molecule and substrate will lead to dissipation of electron and energy, then bring the molecule back to initial equilibrium state. 
The decay time $t_{dec}$ is obtained from the fitting of oscillation amplitude with a exponential function $F(t)=Ae^{-t/t_{dec}}$, the blue line in Fig.\ref{Osci}b. 
The decay time of oscillation is $t_{dec} = 48 fs$, which is much larger than the time-length of the incident laser pulse. 
We also simulate the electronic dynamics excited by laser pulse $t_{cyc} = 1.4fs$ (Fig.\ref{Osci}c), and we can get the same decay time $t_{dec} = 48 fs$ while the corresponding oscillation amplitude is smaller.  
The frequency of oscillation is obtained with Fourier transform, as shown in Fig.\ref{Osci}d.  
The frequencies for various cycle time of pulses $t_{cyc} = (1.0,1.2,1.4,1.6)fs$ are evaluated to be the same value $1.057 fs^{-1}$ $(4.37eV)$, and the cycle times of oscillation are $t_{osci} = 0.946 fs$.   
In the other words, $t_{osci}$ is independent from the stimulated source of laser pulses and exhibits the inherent electronic properties of naphthalene molecule. 
And $t_{osci}$ is the travel time for an electron moving from one side to other side of molecule and reflected back in a round trip.
The travelling velocity $1.04 \times 10^{6} m/s$ is close to the Fermi velocity in graphene.\cite{Trev2008Ab} 
In this way, if $t_{osci}$ could be detected in the experiment, we can also use it to determine the electronic velocity in corresponding molecule. 
When $t_{cyc} = 1.0fs$, the resonance between laser pulse and electronic oscillation would enhance the oscillation amplitude.  
As increase of $t_{cyc}$, the detuning of laser pulses would lead to decrease of oscillation amplitude. 
From the viewpoint of electronic excitation, this oscillation is related to the excitations from HOMO-1 to LUMO and from HOMO to LUMO+1.
The oscillation for excitation from HOMO to LUMO cannot be observed by polarization of pulse in $x$-axis due to the spatial symmetry of HOMO and LUMO, while it can be observed for the polarization in $y$-axis (see supplementary text section II for detail).  

\begin{figure}[htb]
  \includegraphics[width=1.0\columnwidth]{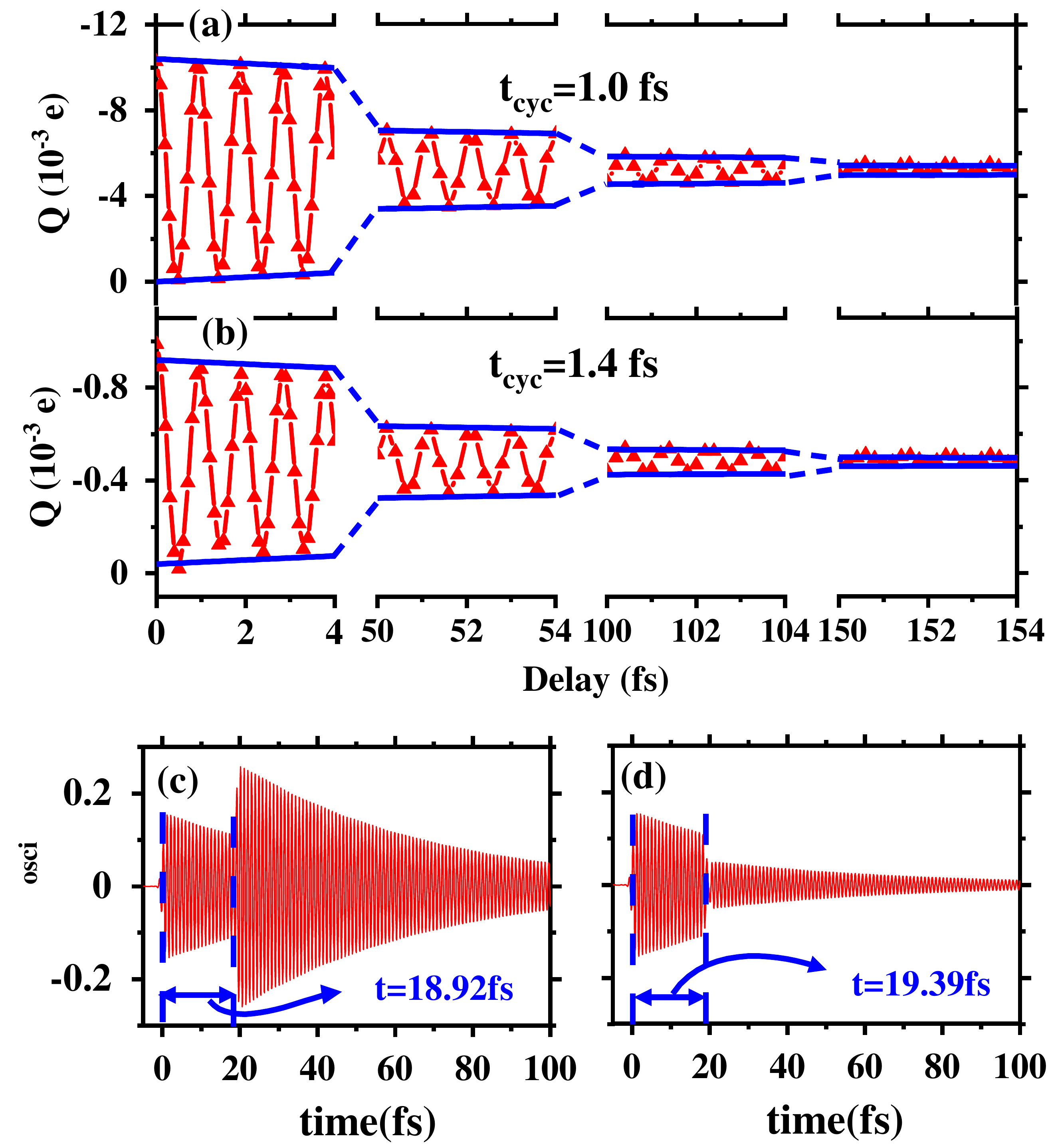}
  \caption{The tunneling charge $Q$ varying with the delay time between pump pulse and probe pulse for $t_{cyc}=1.0fs$ (a) and $t_{cyc}=1.4fs$ (b). The difference of electron density at two sides of molecule $\Delta \rho_{osci}$ excited by pump and probe pulses with delay time $\Delta t = 20 t_{osci}$ (c) and $\Delta t = 20.5 t_{osci}$ (d). }
  \label{pumpProbe}
\end{figure}

To track the electronic oscillation of molecule with tunneling charge $Q$, the pump-probe measurement with laser pulses is used in our simulation.   
The tuning of delay time $\Delta t$ between these two identical pulses can be experimentally realized by a Michelson interferometer, and has been employed to detect similar electronic oscillation inside a Au nanorod.\cite{Garg2020Atto}   
The first pulse (pump pulse) would excite the electronic oscillation in molecule what we attempt to probe using the second pulse (probe pulse).  
Through varying the applied time of probe pulse, Fig.\ref{pumpProbe}a and Fig.\ref{pumpProbe}b plot the simulation results of tunneling charge $Q$ as a function of the delay times $\Delta t$ for laser pulses $t_{cyc} = 1.0fs$ and $t_{cyc} = 1.4 fs$, respectively. 
Both of simulation results show significant oscillating behavior of $Q$ with the same cycle time $0.946 fs$ as electron oscillation $t_{osci}$.    
The maximum value of $Q$ is obtained when delay time $\Delta t$ equals to integral multiple of $t_{osci}$ and minimum value is obtained when $\Delta t$ equals to half-integral multiple of $t_{osci}$. 
This difference of tunnelling charge is due to the interference between the electronic oscillations stimulated by pump pulse and probe pulse. 
When $\Delta t$ equals to integral multiple of $t_{osci}$, such as $\Delta t = 20 t_{osci}$ (Fig.\ref{pumpProbe}c), the constructive interference of electronic oscillations would lead to the maximum value of $Q$. 
In contrast, the destructive interference would lead to the minimum value of $Q$ at $\Delta t = 20.5 t_{osci}$ (Fig.\ref{pumpProbe}d). 
The interference strength should depend on the amplitudes of these two electronic oscillations.  
At the start time of the interference (the applied time of probe pulse), the amplitude of oscillation induced by the pump pulse will decay as the increase of delay time $\Delta t$. 
Consequently, the decrease of interference strength will result in the decay of the oscillating behavior for $Q$.    
And the decay time $t_{dec}^Q$ for this oscillating behavior of $Q$ is got from exponential function fitting: $Q(t)=Q_0 \pm Ae^{-\Delta t/t_{dec}^Q}$ (the blue lines in Fig.\ref{pumpProbe}a and Fig.\ref{pumpProbe}b).  
And this decay time $t_{dec}^Q = 48 fs$ is consistent with the decay time of electronic oscillation $t_{dec}$. 
%
%

%
In summary, using simulations, we have demonstrated the electronic oscillation inside the molecule can be detected by the tunneling microscopy. 
Due to the interference of electronic dynamics excited by pump and probe pulses, the frequency and decay time of oscillation are measured through the varying tunneling charge $Q$ with the tuning of delay time $\Delta t$.      
And this electronic oscillation can be controlled by the frequency and polarization of laser pulse.    
We also show the CEP sensitivity of $Q$ would disappear when the time width of isolated pulses $\sigma$ is approaching the cycle time of pulses $t_{cyc}$.  
And the tunneling charge will linearly increase with the time-length of pulses.  

\acknowledgments

The support from the NSFC (Grant No.\,21803035) and the Natural Science Foundation of Shandong province (Grant No.\,ZR2019BA013) is gratefully acknowledged.
%

\bibliography{bibrefs}

\end{document}